\documentclass[amsmath,twocolumn,showpacs,aps,prl]{revtex4}
\usepackage{bm}
\usepackage{graphics}
\usepackage{graphicx}

\begin{document}

\title{Spin injection dependent metamagnetic transition}
\author{A. A. Zyuzin and A. Yu. Zyuzin}

\affiliation{A.F. Ioffe Physico-Technical Institute of Russian
Academy of Sciences, 194021 St. Petersburg, Russia}


\begin{abstract}
We define the metamagnetic phase transition of itinerant electrons controlled by the spin injection mechanism.
The current flow between a ferromagnetic metal and a metamagnetic metal
produces the non-equilibrium shift of chemical
potential for spin up and spin down electrons that acts as an effective magnetic field driving the
metamagnetic transition.
\end{abstract}
\maketitle

\section{Introduction and main results}
The idea of the spin injection from ferromagneic metal to
paramagnetic metal was first proposed by Aronov
\cite{bib:Aronov2}. In the spin injection process the charge
current flow between the ferromagnetic and paramagnetic metals
produces the non-equilibrium magnetization in the paramagnet. This
magnetization is proportional to the induced chemical potentials
difference of electrons with opposite spins \cite{bib:Aronov2} -
the spin accumulation. Non-equilibrium spin imbalance due to
injection was observed by Johnson and Silsbee \cite{bib:Johnson}.
The theory of spin injection was developed in details in many
works \cite{bib:Son, bib:Johnson2, bib:Valet, bib:Hershfield,
bib:Rashba} and well studied experimentally, see for a review
\cite{bib:Review, bib:Dyakonov}. However, the degree of electron
spin polarization is relatively small at standard spin injection
from ferromagnetic to paramagnetic metal \cite{bib:Jedema1,
bib:Jedema2}. In order to increase the non-equilibrium
polarization it is interesting to look for the possibility of
spin-injection based magnetic transition in metamagnetic metals.
Here we consider the metamagnetic transition of itinerant
electrons induced by the spin injection mechanism. Let us briefly
describe the properties of the metamagnet \cite{bib:Wohlfarth,
bib:Goto}. When the energy splitting of electrons with opposite
spins is smaller than the characteristic energy scale of itinerant
electrons, then magnetic part of the free energy density can be
expanded in powers of magnetization $F(H,M) = aM^2 +bM^4+cM^4-MH$,
where coefficients $a,b,c$ are determined by the energy dependence
of the density of states at the Fermi level, $H$ is the external
magnetic field and $M$ is the magnetization.

\begin{figure}[t] \centering
\includegraphics[width=8cm]{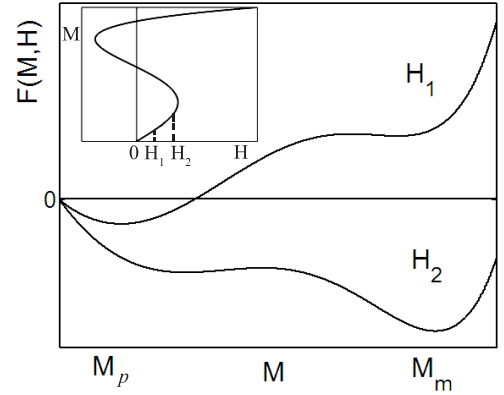}
\caption{Free energy $F(H,M)$ dependence on the magnetization $M$
of the metamagnet shown schematically for different magnetic
fields $H_2>H_1$. The state with higher magnetization has lower
energy at higher magnetic fields. Inset: Dependence of the
magnetization on magnetic field.}\label{fig:1}
\end{figure}

At $b<0$ magnetic part of free energy $F(H=0,M)$ might have
extremum at nonzero $\lvert M\rvert$ as it is shown in Fig.
\ref{fig:1}, which schematically illustrates evolution of free
energy with increasing magnetic field due to contribution of the
term $-MH$. At small magnetic field the state with low
magnetization has lower energy, while at magnetic field larger
than so called metamagnetic field $H_{m}$ the metamagnetic state
acquires lower energy and system undergoes to state with higher
magnetization. Metamagnetic state is induced by external magnetic
field through the first order phase transition
\cite{bib:Wohlfarth, bib:Goto}.

Metamagnetic transition of itinerant electrons might appear
\cite{bib:Wohlfarth, bib:Goto} in strongly enhanced Pauli
paramagnets when the Fermi level is close to peak in electron
density of states. In this case Zeeman splitting increases the
density of states and drives the ferromagnetic transition.

\begin{figure}[t] \centering
\includegraphics[width=8cm]{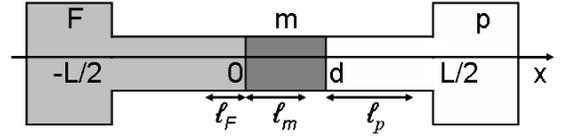}
\caption{Ferromagnetic (-L/2,0) metamagnetic (0,L/2) contact.
Shaded region defines the high magnetization state of the
metamagnet induced by the spin injection from the ferromagnet.
$\ell_{F}, \ell_{m}, \ell_{p}$ are the spin diffusion lengths in
ferromagnet and metamagnet in high and low magnetization
regimes.}\label{fig:2}
\end{figure}

\begin{figure}[t] \centering
\includegraphics[width=8cm]{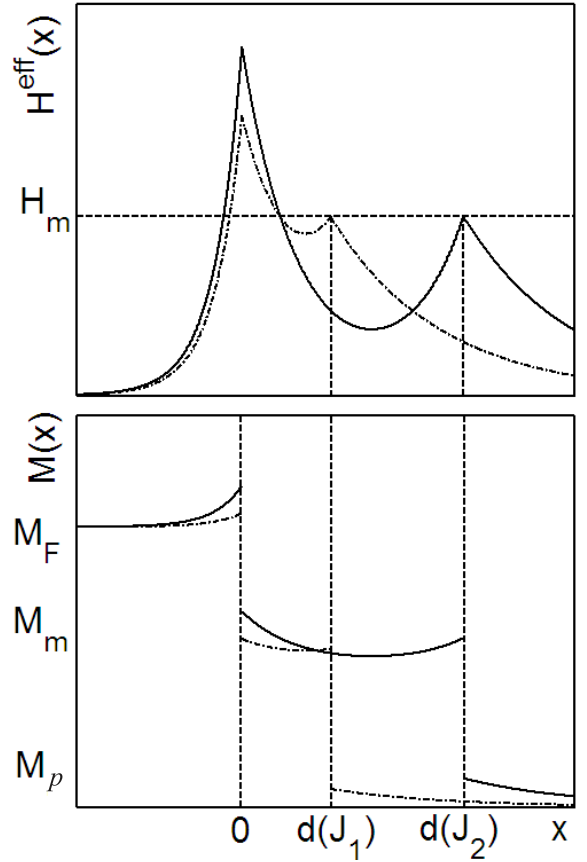}
\caption{Up: Dependence of the effective magnetic field on
coordinate for two values of the current density $|J_2|>|J_1|$.
Effective field decreases in the metamagnet and the phase
transition undergoes at $x=d(J)$ when $H^{\mathrm{eff}} = H_m$.
Down: Magnetization profile, where $M_F, M_m, M_p$ are the
corresponding magnetizations of the ferromagnet, high and low
magnetization states of the metamagnet}\label{fig:3}
\end{figure}

The chemical potentials difference of electrons with opposite
spins is the analog of external magnetic field in the
non-equilibrium case. Characteristic feature of this effective
magnetic field $H^{\textmd{eff}}(x)$ is the spatial non-uniformity
which results in the finite length of the metamagnetic state. In
the ferromagnet - metamagnetic metal contact spin accumulation and
therefore effective magnetic field is generated in the region of
the order of spin relaxation length at the vicinity of contact
with ferromagnet and at the domain wall between metamagnetic -
paramagnetic states. We assume that the domain wall thickness is
much smaller than the spin relaxation length.
Schematically ferromagnetic metal - metamagnetic state contact is
shown in Fig. \ref{fig:2}. Metamagnetic state is located at
$0<x<d$. Metamagnetic phase emerges at electric currents such that
the effective magnetic field is $H^{\mathrm{eff}}(x=0)\geq H_{m}$. If
$d$ is of order or larger than the spin relaxation length then
effective field $H^{\textmd{eff}}(x)$ can be estimated as a sum
$H^{\textmd{eff}}(x)=H_{F-m}^{\textmd{eff}}(x)+H_{m-p}^{\textmd{eff}}(x)$
of the fields due to spin accumulation at boundary $x=0$
\begin{equation}\label{F-m}
H_{F-m}^{\textmd{eff}}(x) =
\frac{eJ}{g\mu_B}\frac{2R_FR_m}{R_F+R_m}[\Pi_F-\Pi_m]e^{-x/\ell_{m}}
\end{equation}
and effective field due to spin accumulation at domain wall
$x=d$
\begin{equation}\label{m-p}
H_{m-p}^{\textmd{eff}}(x) =
\frac{eJ}{g\mu_B}\frac{2R_mR_p}{R_m+R_p}\Pi_{m}
e^{-(d-x)/\ell_{m}}
\end{equation}
This case is shown by the solid line in Fig. \ref{fig:3}. In
expressions (\ref{F-m}) and (\ref{m-p}) $J$ is the current
density, $e$ is the electron charge, $\mu_B = |e|\hbar/2mc$ is the
Bohr magneton and $g=2$ for electrons,
\begin{eqnarray}\label{pi}
\Pi_{F,m} = \frac{\sigma_{\uparrow F,m}-\sigma_{\downarrow
F,m}}{\sigma_{\uparrow F,m}+\sigma_{\downarrow F,m}}
\end{eqnarray}
is proportional to the current polarizations, where
$\sigma_{\alpha} = e^2 D_{\alpha} \nu_{\alpha}$ are the
corresponding conductivities in the ferromagnetic, metamagnetic
and paramagnetic states for electrons with spin $\alpha$,
$D_{\alpha}$ is the diffusion coefficient, $\nu_{\alpha}$ is the
density of states at the Fermi level,
\begin{equation}\label{resist}
R_{F,m}=\ell_{F,m}\frac{\sigma_{\uparrow F,m}+\sigma_{\downarrow
F,m}}{4\sigma_{\uparrow F,m}\sigma_{\downarrow F,m}} , R_p =
\frac{\ell_{p}}{\sigma_{p}}
\end{equation}
are the effective resistances and the spin relaxation lengths are
defined as $\ell = \sqrt{\overline{D}t_s}$, where in each state
$\overline{D} = (D_{\uparrow}\sigma_{\downarrow} +
D_{\uparrow}\sigma_{\downarrow})/(\sigma_{\uparrow}+\sigma_{\downarrow})$
and $t_s$ is spin relaxation time.

In case of small thickness of domain wall transition between metamagnetic and paramagnetic states takes
place at $x=d$ when
\begin{equation}\label{main}
H^{\textmd{eff}}(d)=H_{m}
\end{equation}
as shown in Fig. \ref{fig:3}. Taking the sum of expressions
(\ref{F-m}) and (\ref{m-p}) we estimate the corresponding length
of the metamagnetic region at $d\geq \ell_{m}$ as
\begin{equation}\label{length}
d \sim \ell_{m}\ln\left|\frac{R_F[\Pi_F -\Pi_m
]/(R_m+R_F)}{g\mu_BH_m/eJ2R_m -R_p\Pi_m/(R_m+R_p)} \right|
\end{equation}
At large electrical current when $g\mu_BH_m/eJ2R_m \rightarrow
R_p\Pi_m/(R_m+R_p)$, according to expression (\ref{length}) the
length of metamagnetic region $d\rightarrow\infty$. Threshold
current density dependence of $d$ occurs because of spin
accumulation generation at domain wall between metamagnetic and
paramagnetic states. However, since the effective field in most
part of the region is below $H_m$, the energy of stationary
metamagnetic state at large $d$ becomes smaller than the energy of
paramagnetic state. We propose that, system must undergo to
paramagnetic state at large values of current density. More
detailed discussion of the transition is given below.
\section{Electrical spin injection}
Consider the spin injection process from the ferromagnetic metal
to metamagnetic metal. We focus on the spin and charge transport
in the presence of the spin-orbit coupling and the short-range
exchange electron-electron interactions. We assume the vector of
the non-equilibrium magnetization in the metamagnetic metal to be
parallel to the vector of the magnetization in the ferromagnet.
The Green's function in Keldysh technique in the matrix form
\begin{equation}\nonumber
\underline{\hat{G}}=\left(\begin{array}{clcr}
\hat{G}^R&\hat{G}^K\\
0&\hat{G}^A
\end{array}\right)
\end{equation}
is given by retarded $\hat{G}^R(\mathrm{x},\mathrm{x}')$, advanced
$\hat{G}^A(\mathrm{x},\mathrm{x}')$ and Keldysh function
$\hat{G}^K(\mathrm{x},\mathrm{x}')$, where
$\mathrm{x}=(\mathbf{r}, t)$ denote position and time arguments,
hat $\left( \hat{} \right)$ stands for the matrix in spin space.
Further we will consider the stationary regime in which function
$\underline{\hat{G}}$ satisfies the equation
\begin{eqnarray}\nonumber
&[(\omega+\frac{1}{2m}\nabla^2_{\mathbf{r}} + \mu -U(\mathbf{r})
-e\phi(\mathbf{r}) )\hat{1}-&\\ \nonumber &-
\hat{U}_{so}(\mathbf{r})+\hat{\varepsilon}(\mathbf{r})]
\underline{\hat{G}}(\mathbf{r},\mathbf{r}',\omega)
=\hat{1}\delta(\mathbf{r}-\mathbf{r}')&
\end{eqnarray}
where $\phi(\mathbf{r})$ is the electrostatic potential,
$U(\mathbf{r})$ is the random potential of the impurities assumed
to be Gaussian distributed with $\langle U(\mathbf{r})\rangle=0$,
$\langle
U(\mathbf{r})U(\mathbf{r}')\rangle=2\pi\nu\tau\delta(\mathbf{r}-\mathbf{r}')$,
$\tau$ is the mean free time and $\nu=(\nu_\uparrow +
\nu_\downarrow)/2$ is the density of states. Spin-orbit
interactions of electrons with impurities is described by the
potential
$\hat{U}_{so}(\mathbf{r})=i\gamma\mathbf{\hat{\sigma}}(\mathbf{\nabla}
U(\mathbf{r})\times\mathbf{\nabla})$, where $\gamma$ is the
spin-orbit coupling constant and $\hat{\sigma}$ is the Pauli
matrix \cite{bib:Altshuler}. The contribution of the short-range
electron-electron exchange interactions to the spin splitting is
described by the term $\hat{\varepsilon}(\mathbf{r})$
\begin{equation}\nonumber
\varepsilon_{\alpha}(\mathbf{r})=\frac{-i\lambda}{2}\int\frac{d\mathbf{p}d\omega}{(2\pi)^4}G^{K}_{-\alpha}(\mathbf{r},
\mathbf{p},\omega)
\end{equation}
where $\lambda$ is the electron coupling constant and it is
convenient to apply the Fourier transformation with respect to the
relative coordinates $\mathbf{r}_1 = \mathbf{r}-\mathbf{r}'$ as
\begin{equation}\nonumber
\underline{\hat{G}}(\mathbf{R},\mathbf{p},\omega) =\int
d^3\mathbf{r}_1 \underline{\hat{G}}(\mathbf{R}+\mathbf{r}_1/2,
\mathbf{R}-\mathbf{r}_1/2) e^{-i\mathbf{pr_1}}
\end{equation}
in which $\mathbf{R}=(\mathbf{r}+\mathbf{r}')/2$. The retarded and
advanced Green functions $\hat{G}^R$ and $\hat{G}^A$ averaged over
disorder in the $\mu\tau\gg1$ approximation are diagonal in spin
and are given by
\begin{equation}\label{green}
\hat{G}^{R,A}(\mathbf{R},\mathbf{p},\omega) =
\frac{1}{\omega-\xi_\mathbf{p}+\hat{\varepsilon}(\mathbf{R})\pm
i\hat{\gamma}}
\end{equation}
where $\xi_\mathbf{p}$ is the electron dispersion,
$2\gamma_{\alpha} = \tau_{\alpha}^{-1}+
(t^{-1}_{\alpha}-t^{-1}_{-\alpha})/2$ and
$\tau_{\alpha}^{-1}=\tau_{0\alpha}^{-1}+\tau_{s\alpha}^{-1}$ is
the inverse scattering times due to disorder and spin-orbit
interactions for the spin state $\alpha$,
$t^{-1}_{s\alpha}=4/3\tau_{s\alpha}$ \cite{bib:Altshuler}. We
assume that the momentum relaxation time $\tau_{0\alpha}$ is
smaller than the time $t_{s \alpha}$ corresponding to the spin
flip process. Let us note, that we are considering the metamagnet
when the exchange energy is the coordinate dependent function. In
deriving the equation for the function $\hat{G}^K$ we assume the
limit when the exchange energy is small compared to the Fermi
energy $\mid\varepsilon_\downarrow -\varepsilon_\uparrow \mid/\mu
< 1$. In this limit the equation for the function $\hat{G}^K$
yields
\begin{eqnarray}\nonumber
&\mathbf{v}(\mathbf{\nabla}_{\mathbf{R}} +
[\mathbf{\nabla}_{\textmd{R}}\varepsilon_{\alpha} +
e\mathbf{E}]\partial_{\varepsilon_p})G^{K}_{\alpha}=-\left(
\frac{1}{\tau_{\alpha}} -\frac{1}{ t_{\alpha s}}+ \frac{1}{
t_{-\alpha s}}\right)G^{K}_{\alpha}&\\ \nonumber&
+\left(\frac{F_{\alpha}}{\tau_{\alpha}} -
\frac{F_{\alpha}-F_{-\alpha}}{2t_{\alpha s
}}\right)[G^{R}_{\alpha}-G^{A}_{\alpha}]&
\end{eqnarray}
here $\mathbf{E}=-\nabla\phi(\mathbf{r})$ is the electric field
and we denote the coordinate and frequency dependent function
\begin{equation}\nonumber
F_{\alpha}(\mathbf{R},\omega)=\frac{i}{2\pi\nu_{\alpha}}\int\frac{d\mathbf{p}}{(2\pi)^3}G^{K}_{\alpha}(\mathbf{R},\mathbf{p},\omega)
\end{equation}
In the diffusion approximation for the function
$F_{\alpha}(\mathbf{R},\omega)$ one obtains the equation
\begin{equation}\label{diff}
\mathbf{\nabla}\sigma_{\alpha}\mathbf{\nabla}F_{\alpha}(\mathbf{R},\omega)
= \frac{\nu_{\alpha}
\nu_{-\alpha}}{2\nu}\frac{F_{\alpha}(\mathbf{R},\omega)-F_{-\alpha}(\mathbf{R},\omega)}{t_{s}}
\end{equation}
where $\sigma_{\alpha} = e^2 \nu_{\alpha} D_{\alpha}$ is the
conductivity, $D_{\alpha} = v^2_{\alpha}\tau_{\alpha}/3$ is the
diffusion coefficient and the density of states are the space
dependent functions.

Let us consider the system when functions in Eq. (\ref{diff})
depend on one spatial coordinate $(x)$ only. Consider the
interface between a ferromagnetic metal that occupies the region
$(x<0)$ and a metamagnetic metal $(x>0)$. We assume that the
lengths of the ferromagnetic and metamagnetic regions $L/2$ are
much larger than the corresponding spin diffusion lengths. We also
assume the external reservoirs of the sample at $x=\pm L/2$ to be
in the spin equilibrium state. The electric field in the system is
treated through the boundary conditions
\begin{eqnarray}\label{boundary}\nonumber
F_{\alpha}(-L/2,\omega) = f(\omega-eV/2)
\\ F_{\alpha}(L/2,\omega) = f(\omega+eV/2)
\end{eqnarray}
where $f(\omega)=\tanh(\omega/2T)$ and $V=E/L$ is the potential
difference across the structure. The solution of the Eq.
(\ref{diff}) is the continuous function at the interface $x=0$
\begin{equation}\label{con1}
F_{\alpha}(0-,\omega)=F_{\alpha}(0+,\omega)
\end{equation}
while the derivatives satisfy
\begin{equation}\label{con2}
\sigma_{\alpha}\frac{\partial F_{\alpha}(x,\omega)}{\partial x}|
_{x=0-} = \sigma_{\alpha}\frac{\partial
F_{\alpha}(x,\omega)}{\partial x}|_{x=0+}
\end{equation}
describing the continuity of the current density at the interface.
The current density carried by spin up and spin down electrons is
given as
\begin{equation}\nonumber
J_{\alpha}(x)= \frac{1}{2e}\int \sigma_{\alpha}\frac{\partial
F_{\alpha}(x,\omega)}{\partial x}d\omega
\end{equation}
We solve Eq. (\ref{diff}) assuming the boundaries
ferromagnet-paramagnet, ferromagnet-metamagnet and
metamagnet-paramagnet independently. This approximation is valid
in the limit when the length of the metamagnet $d>\ell_m$. Taking
into account the length of the system to be larger than the
corresponding spin-diffusion lengths we solve the diffusion
equation in the region $x>0$ with boundary (\ref{boundary}) and
continuity (\ref{con1}), (\ref{con2}) conditions.
\begin{eqnarray}\label{solution}\nonumber
&&F_{ \uparrow, \downarrow| p, m}(x,\omega)= f(\omega+eV/2)
+A_{p,m}[(x-L/2) \pm\\
&&\pm 2\sigma_{\downarrow, \uparrow| p, m}(\Pi_F-\Pi_{p,
m})\frac{R_F R_{p, m}}{(R_F+R_{p, m})}e^{-x/\ell_{p,m}}]
\end{eqnarray}
where $p, m$ denotes low and high magnetization regimes of the
metamagnet and $F$ stands for the ferromagnet, coefficient
\begin{eqnarray}\nonumber
A_{p,m} = \frac{2(\sigma_{\uparrow F}+ \sigma_{\downarrow
F})[f(\omega+eV/2)-f(\omega-eV/2)]}{ [(\sigma_{\uparrow F}+
\sigma_{\downarrow F})+(\sigma_{\uparrow |p, m}+
\sigma_{\downarrow |p, m})]L }
\end{eqnarray}
is connected with the current density as
\begin{eqnarray}\nonumber
J=J_{\uparrow}(x)+J_{\downarrow}(x)=\frac{1}{2e}\int
[\sigma_{\uparrow |p, m}+\sigma_{\downarrow |p, m}]A_{p,m}d\omega
\end{eqnarray}
The conductivity spin polarization and resistivities in the
ferromagnet and metamagnet are defined by expressions (\ref{pi})
and (\ref{resist}). Note, that in the low magnetization regime of
the metamagnet  $\sigma_{\uparrow p} = \sigma_{\downarrow p} =
\sigma_{p}/2$, $D_{\uparrow p} = D_{\downarrow p} = D_{p}$.
Solution (\ref{solution}) has to be supplemented by the local
neutrality condition which self-consistently determines the
electric potential in the sample. The spin injection process does
not change concentration of electrons
\begin{equation}\label{neitrality}
N = \frac{1}{2}\int
[\nu_{\uparrow}F_{\uparrow}(x,\omega)+\nu_{\downarrow}F_{\downarrow}(x,\omega)]d\omega
\end{equation}

\section{Paramagnetic state} Low magnetization state of the
metamagnet can be studied by solving Eq. (\ref{diff}) assuming
contact between ferromagnetic metal and paramagnetic metal at
$x=0$. Effective field due to spin accumulation is
\begin{equation}\nonumber
H_p^{\mathrm{eff}}(x) = \frac{eJ}{g\mu_B} \frac{2R_F
R_p}{R_F+R_p}\Pi_Fe^{-x/\ell_{p}}
\end{equation}
and magnetization at $x>0$ is
\begin{equation}\label{para}
M_{p}(x) =\frac{(g\mu_B)^2
\nu_p}{1-\lambda\nu_p}H_{p}^{\mathrm{eff}}(x)
\end{equation}
Effective magnetic field produced by the spin accumulation in the
ferromagnetic metal at $x<0$ is
\begin{equation}\label{FerroFieldF-N}
H_{Fp}^{\mathrm{eff}} (x) = \frac{eJ}{g\mu_B} \frac{2R_F
R_{p}}{R_F+R_{p}}\Pi_{F}e^{x/\ell_{F}}
\end{equation}
here expressions for resistances $R_{F}$ and $R_{p}$ are given by
Eq. (\ref{resist}).
\section{Metamagnetic transition}
The self-consistency equation for the magnetization density $M(x)$ in the sample is
defined as
\begin{eqnarray}\label{magnetization}\nonumber
&M(x) =g\mu_B[\varepsilon_{\downarrow}(x) -
\varepsilon_{\uparrow}(x)]/\lambda =&\\& = -\frac{g\mu_B}{2}\int
[\nu_{\uparrow}F_{\uparrow}(x,\omega)-\nu_{\downarrow}F_{\downarrow}(x,\omega)]
d\omega&
\end{eqnarray}
In the case of equilibrium  metamagnetic metal, Eq.
(\ref{magnetization}) has two solutions even without the external
magnetic field, corresponding to two minima of free energy, see
inset in Fig. (\ref{fig:1}). Transition between these solutions
takes place when magnetic field is equal to $H_{m}$. One could
verify that in linear on $V$ response spin dependent part of
expression (\ref{solution}) enters Eqs. (\ref{neitrality},
\ref{magnetization}) as magnetic field.

The procedure of finding solutions is following. We assume that
there is metamagnetic state in the system at $0<x<d$. Then we
solve Eq. (\ref{diff}) for the spin accumulation at two boundaries
and self consistently determine the value of $d$ from Eq.
(\ref{main}).

To obtain Eq. (\ref{main}) we need to consider transition in more
detail. Near transition between metamagnetic and paramagnetic
states we need to include the spatial derivatives of magnetization
into consideration, so
\begin{equation}\label{x}\nonumber
-K\frac{d^2}{dx^2}M+\frac{\delta F(H_{m}^{\mathrm{eff}} (x),M)}{\delta
M}=0
\end{equation}
Here $K$ is positive constant. Let we have solution $M_{w}(x-d)$,
describing transition between metamagnetic and paramagnetic states
at point $x=d$ in uniform magnetic field. It is solution of
equation
\begin{equation}\label{y}\nonumber
-K\frac{d^2}{dx^2}M_{w}+\frac{\delta F(H_{m},M_{w})}{\delta M_{w}}=0
\end{equation}
Assuming small difference $H_{m}^{\mathrm{eff}} (x)-H_{m}$ at
$x\approx d$ and substituting $M=M_{w}(x-d)+\delta M$, we obtain
\begin{equation}\label{delta-M}\nonumber
-K\frac{d^2}{dx^2}\delta M+\frac{1}{2}\frac{\delta ^2 F(H_{m},M)}{\delta
M^2}|_{M=M_{w}}\delta M=H_{m}^{\mathrm{eff}} (x)-H_{m}
\end{equation}
Solution of this equation exists if
\begin{equation}\label{condition}
\int dx\Psi_{0}(x)(H_{m}^{\mathrm{eff}} (x)-H_{m})=0
\end{equation}
where $\Psi_{0}(x)$ is eigenfunction, corresponding to zero
$E_{0}=0$ mode of equation
\begin{equation}\label{z}\nonumber
-K\frac{d^2}{dx^2}\Psi_{k}+\frac{1}{2}\frac{\delta ^2
F(H_{m},M)}{\delta M^2}|_{M=M_{w}}\Psi_{k}=E_{k}\Psi_{k}
\end{equation}
$\Psi_{0}(x)$ has no zeros and is localized near $x=d$ in region
of order of domain wall thickness. This mode describes small
translation of $M$, so in the uniform field one has $E_{0}=0$. In
the case when $\ell_{m,p}$ are much larger than domain wall
thickness from condition (\ref{condition}) one obtains Eq.
(\ref{main}).

\subsection{Metamagnetic state}
Let us discuss different realizations of spin accumulation.

1). Consider the case when $\Pi_F-\Pi_m$ and $\Pi_m$ have the same
sign. Effective fields of both contacts have same sign too. The
estimation for the length of  metamagnetic region $d$ in the limit
$d\geq\ell_{m}$ is given by expression (\ref{length}). $d$
diverges at some threshold electrical current density.

2). Let $\Pi_F-\Pi_m$ and $\Pi_m$ have opposite signs. Thus,
effective fields of both contacts have different signs too.
Analysis shows that solution of Eq.(\ref{main}) with finite $d$
exist at
$|H_{F-m}^{\textmd{eff}}(0)|>|H_{m-p}^{\textmd{eff}}(d)|$. With
incresing electrical current density $d$ stays finite.

In metamagnetic region the magnetization is
\begin{equation}\label{Magnet-meta}\nonumber
M_m (x) = M_{m}^{0} +\frac{(g\mu_B)^2\nu_m}{1-\lambda \nu_m}
H_m^{\mathrm{eff}}(x)
\end{equation}
Here $M_{m}^{0}$ is the magnetization of metamagnetic state,
calculated at zero magnetic field and $\nu_m = 2\nu_{\uparrow
m}\nu_{\downarrow m}/(\nu_{\uparrow m}+\nu_{\downarrow m})$.

Spin accumulation appears also in the ferromagnetic metal at $x<0$
as
\begin{equation}\label{FerroFieldF-m}
H_{Fm}^{\mathrm{eff}} (x) = \frac{eJ}{g\mu_B} \frac{2R_F
R_{m}}{R_F+R_{m}}[\Pi_F - \Pi_{m}]e^{x/\ell_{F}}
\end{equation}
here expressions for $R_{F}$ and $R_{m}$ are given by
(\ref{resist}).

\subsection{Free energy criterium}
We propose that the realization of metamagnetic state must be
energetically favorable over realization of the paramagnetic
state. In the linear on the applied current regime the magnetic
part of free energy in the case of paramagnetic state realization
is
\begin{equation}\nonumber
\delta\mathcal{F}_{\textmd{Fp}} = - M_F
\int_{-L/2}^{0}H^{\mathrm{eff}}_{Fp}(x) dx
\end{equation}
where $M_F$ is the magnetization of ferromagnetic contact. In the
case of metamagnetic state realization it is
\begin{eqnarray}\nonumber
&\delta\mathcal{F}_{\textmd{Fmp}} = - M_F \int_{-L/2}^{0}
H^{\mathrm{eff}}_{Fm}(x) dx -&\\ \nonumber&- M^{0}_m
\int_{0}^{d}[H^{\mathrm{eff}}_{m}(x)-H_{m}] dx+F_{S}&
\end{eqnarray}
Effective magnetic fields in ferromagnetic region are given by
expressions (\ref{FerroFieldF-N}) and (\ref{FerroFieldF-m}).
$F_{S}$ is the energy, associated with domain wall and boundary
$F-m$. While domain wall energy is positive, the sign of $F-m$
boundary energy depends on relative directions of magnetizations
in the ferromagnet and metamagnet. Estimation of $F_{S}$ depends
on details that are beyond the scope of the paper.

From criterium $\delta \mathcal{F}_{\textmd{Fp}}-\delta
\mathcal{F}_{\textmd{Fmp}}\geq 0$ for realization of the
metamagnetic transition one can estimate the threshold value of
current density. In the limiting case $R_F>R_m>R_p$, which assumes
the contribution of the boundary $m-p$ and $F-p$ to the free
energy is smaller than the corresponding contribution from the
$F-m$ interface, one can estimate as
\begin{eqnarray}\nonumber
J_{thr}\approx\frac{g\mu_B}{e}\frac{
H_m/2R_m}{(\Pi_F-\Pi_m)(1+\ell_F M_F/\ell_m
M^0_m)}\frac{d(J_{thr})}{\ell_m}
\end{eqnarray}

\subsection{Ferromagnet-metamagnet-ferromagnet structure}
Let us briefly discuss the spin injected metamagnetic state in
system with metamagnetic metal placed between two ferromagnetic
contacts with opposite directions of magnetizations. In this case
$\delta\mathcal{F}_{\textmd{Fp}}=0$, because of cancelation of
contributions in ferromagnets with opposite magnetizations. In
metamagnetic state
\begin{eqnarray}\nonumber
&\delta\mathcal{F}_{\textmd{FmF}} = - M^{0}_m
\int_{0}^{d}[H^{\mathrm{eff}}_{m}(x)-H_{m}] dx&
\end{eqnarray}
Both ferromagnets contribute equally to the effective field. At
$d\geq \ell_{m}$ using expression (\ref{F-m}) we obtain the
threshold value of electrical current density at which
$\delta\mathcal{F}_{\textmd{FmF}}\leq 0$ as
\begin{equation}\nonumber
J_{thr}=\frac{g\mu_B}{e}H_{m}\frac{R_F+R_m}{4R_FR_m[\Pi_F-\Pi_m]}\frac{d}{\ell_{m}}
\end{equation}
Note, that the expression for $J_{thr}$ for the $F-m$ contact in
the limit discussed in the previous section is similar to the
$F-m-F$ contact. Also note, that the transition to paramagnetic
state with increasing current is absent.
\section{conclusions}
To conclude, we have studied the metamagnetic transition of
itinerant electrons in the metamagnet under the spin injection
from the ferromagnetic metal. Spin injection produces the
non-equilibrium effective magnetic field in metamagnet which
drives the transition. We have calculated the effective magnetic
fields and  electrical currents required for the metamagnetic
transition. We have shown that the length of metamagnetic state
has threshold dependence on electrical current due to the
effective magnetic field self generated at domain wall.

Typical values of the spin accumulation in metals are in the $\mu
\mathrm{eV}$ range \cite{bib:Fert, bib:Zaffalon}, which
corresponds to the effective magnetic fields in tenth
$\mathrm{mT}$ range at reasonably high current density. Metallic
metamagnets with metamagnetic field in tesla's range are well
known \cite{bib:Gto}. Applying external magnetic field one can
easily bring such system close to the transition.

Well studied $\mathrm{YCo_{2}}$ in crystal form undergoes the
metamagnetic transition at $H_{m}=70T$ \cite{bib:Goto}, while in
the nanoscale structured form it is a weak ferromagnet
\cite{bib:Narayana}, suggesting the possibility of metamagnetic
field reducing at proper technology. Other possibility is to study
the system with temperature induced metamagnetic transition
\cite{bib:Markosyan}. Unfortunately, spin relaxation length, the
main parameter that governs the magnitude as well as the spatial
distribution of effective field, is not known in such systems.

\section{Acknowledgments}
We thank V.I. Kozub and A.T. Burkov for valuable discussion. We
are grateful for the financial support of Federal Program under
Grant No. 2009-1.5-508-008-012 and RFFI under Grant No.
10-02-00681-A.

\end{document}